\documentclass[twocolumn,prb,showpacs]{revtex4}
\usepackage{graphicx}
\usepackage{bm}

\begin{document}
\title{Exact low-temperature behavior of kagom\'e antiferromagnet
at high fields}

\author{M. E. Zhitomirsky}
\affiliation{
Commissariat \`a l'Energie Atomique, DSM/DRFMC/SPSMS, 38054 Grenoble, 
France}
\author{Hirokazu Tsunetsugu}
\affiliation{
Yukawa Institute for Theoretical Physics, Kyoto University, 
Kyoto 606-8502, Japan}

\date{May 25, 2004}

\begin{abstract}
Low-energy degrees of freedom of a spin-1/2 kagom\'e antiferromagnet in 
the vicinity of the saturation field are mapped to a hard-hexagon model
on a triangular lattice. The latter model is exactly solvable.
The presented mapping allows to obtain quantitative description
of the magnetothermodynamics of a quantum kagom\'e antiferromagnet
up to exponentially small corrections as well as predict the critical
behavior for the transition into a magnon crystal state.
Analogous mapping is presented for the sawtooth chain, which
is mapped onto a model of classical hard dimers on a chain.
\end{abstract}

\maketitle

\begin{figure}[t]
\begin{center}
\includegraphics[width=0.85\columnwidth]{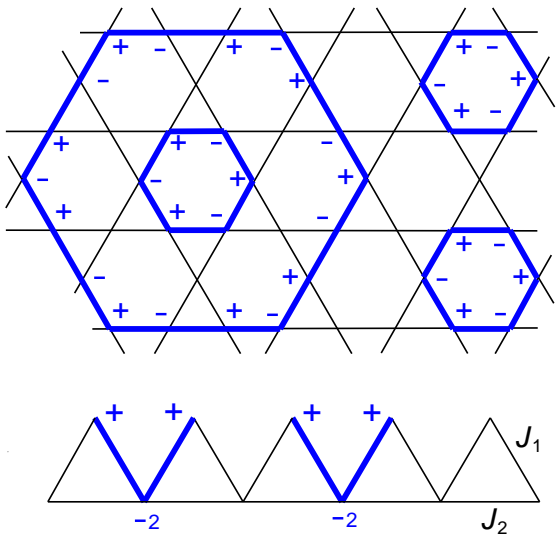}
\end{center}
\caption{(color online)
Kagom\'e lattice (top) and the sawtooth chain (bottom) with
localized magnons.
Amplitudes and phases of spin-down states 
are indicated near each site.
\label{geometry}} 
\end{figure}

In the past few years there have been increased attention to
the behavior of frustrated magnetic materials at high fields. The 
interest is stimulated by a search for new fundamental effects 
such as the magnetization plateaus,\cite{kageyama,miyahara,%
momoi,hiroi,ogata02,moessner03,hida,mzh02,cabra} the magnetization 
jumps,\cite{schulenburg,richter,schnack,schmidt} new 
complicated magnetic phase diagrams,\cite{mzh02,ramirez02}
and new types of the critical behavior\cite{jackeli} as well 
as by a possible technological application of frustrated magnets
in the adiabatic demagnetization refrigerators due to their unique
magnetocaloric properties.\cite{mzh03,sosin} A zero-temperature 
behavior of geometrically frustrated antiferromagnets in the vicinity 
of the saturation field $H_c$ has been recently discussed by 
Schulenburg and co-workers.\cite{schulenburg,richter,schnack,schmidt}  
Above $H_c$ in the fully polarized phase, these models have a flat 
branch of magnons with energy $(H-H_c)$. Such dispersionless 
excitations correspond in real space to localized spin flips, which 
do not interact with each other unless touched. Upon decreasing 
magnetic field through $H_c$, the energy of a single excitation 
becomes negative and magnons condense into a close-packed crystal 
structure. The present work is devoted to theoretical investigation 
of the low-temperature thermodynamics of magnon crystals.
Specifically, we consider spin-1/2 Heisenberg  models 
on a kagom\'e lattice and a sawtooth (or $\Delta$) chain,
see Fig.~\ref{geometry}. While the former lattice
is one of the best known examples of geometric frustration, which 
is realized, e.g., in SrCr$_{9p}$Ga$_{12-9p}$O$_{19}$,\cite{scgo}
the latter spin model is applicable to magnetic delafossite
YCuO$_{2.5}$.\cite{delafossite} We show that the thermodynamic
potential in the vicinity of $H_c$  for the two quantum models can
be calculated exactly up to corrections, 
which are exponentially small at low temperatures.

We consider nearest-neighbor spin-1/2 Heisenberg models in an 
external field
\begin{equation}
\hat{\cal H} = \sum_{\langle ij\rangle} J_{ij} 
{\bf S}_i\cdot {\bf S}_j
- {\bf H}\cdot \sum_i {\bf S}_i \ .
\label{Hamiltonian}
\end{equation}
For a kagom\'e lattice all nearest-neighbor bonds have the same
strength $J_{ij}\equiv 1$. The spectrum of single-magnon 
excitations in the fully polarized phase is
\begin{equation}
\omega_{1{\bf k}}=H-3 \ ,\ \omega_{2,3{\bf k}}=H - 
{\textstyle \frac{3}{2}} \pm
{\textstyle \frac{1}{2}}\sqrt{3(1+2\gamma_{\bf k})} \ , 
\label{wkag}
\end{equation}
where $\gamma_{\bf k} = \frac{1}{6} \sum_\delta 
e^{i{\bf k}\cdot\delta}$ is a sum over nearest-neighbor 
sites on a triangular Bravais lattice.
At the saturation field $H_c=3$ the energy of excitations 
from the lowest dispersionless branch vanishes. 
The sawtooth chain exhibits a similar behavior for the special
ratio of the two coupling constants $J_2=\frac{1}{2}J_1$.
In the following we always assume the above choice of 
the parameters with $J_1\equiv 1$. Above $H_c=2$ the sawtooth 
chain antiferromagnet has two branches of single-magnon 
excitations:
\begin{equation}
\omega_{1k}=H-2\ ,\ \ \omega_{2k}=H - {{\textstyle \frac{1}{2}}}
(1-\cos k) \ .
\label{wsaw}
\end{equation}
Excitations in the dispersionless branches correspond to localized 
states. Simplest examples of such states are spin 
flips trapped on hexagon voids of a kagom\'e lattice or in the 
valleys of the sawtooth chain, see Fig.~\ref{geometry}. 
Their wave functions are given by 
\begin{eqnarray}
|\varphi_i\rangle & = & {\textstyle \frac{1}{\sqrt{6}}}
\sum_{n=1}^6 (-1)^{n-1}S^-_{ni}|{\rm FM}\rangle \  , \nonumber \\
|\varphi_i\rangle & = & {\textstyle \frac{1}{\sqrt{6}}}
(S^-_{1i} - 2 S^-_{2i} +S^-_{3i}) |{\rm FM}\rangle \ ,
\label{local}
\end{eqnarray}
for the kagom\'e and the sawtooth models, respectively, 
$|{\rm FM}\rangle$ being the ferromagnetic state. Other 
more extended localized states can be represented as linear 
combinations of the above elementary states. Moreover,
for a kagom\'e lattice with $N$ sites, there is 
the same number of hexagons $\frac{1}{3}N$ as the number of states
in the flat branch. Therefore, the localized 
states on the hexagons states can be used as a real-space
basis for the lowest branch in Eq.~(\ref{wkag}). Similar 
correspondence exists between the $\frac{1}{2}N$ states in the dispersionless
branch Eq.~(\ref{wsaw}) and the localized states in the 
valleys of the sawtooth chain.

Localized one-magnon states allow to construct a class of exact
eigenstates in every $n$-magnon subsector with 
$n\leq N_{\rm max}$ by putting localized spin flips on 
isolated hexagons or valleys. The highest possible density 
of independent localized magnons on the kagom\'e lattice 
is one-third of the total number of hexagons or 
$N_{\rm max}=\frac{1}{9}N$. Such a magnon crystal state is three-fold
degenerate and breaks the translational symmetry. A magnon  
crystal for the sawtooth chain has $N_{\rm max}=\frac{1}{4}N$ 
and is two-fold degenerate. Since isolated localized 
magnons do not interact with each other, it is natural 
that they correspond to the lowest energy states in every 
$n$-magnon subsector ($n\leq N_{\rm max}$). 
\cite{schulenburg,richter,schnack,schmidt} Our idea is that 
only isolated localized  magnons contribute to the low
temperature thermodynamics. In order to prove this 
we have to show that (i) there are no other low energy 
states and that (ii) isolated localized magnon states
are separated by a finite gap from the higher-energy states
in the same $S^z$-subsector. In the following we first prove
the above two points and then show that localized magnons 
are mapped to statistical mechanics models 
of hard classical objects, hexagons or dimers,
which allow exact solution for the thermodynamic properties.

Numerical evidences and analytical arguments
that isolated localized magnon states
are the lowest energy states in every $n$-magnon sector
have been presented before. 
\cite{schulenburg,richter,schnack,schmidt}
Below we give a shortened
proof for the spin-1/2 kagom\'e antiferromagnet,
which is also valid for the checkerboard
and the pyrochlore lattice antiferromagnets.
The Heisenberg Hamiltonian (\ref{Hamiltonian})
is split in a standard way into
the Ising part $\hat{\cal H}^{zz}$, which includes
all $S_i^zS_j^z$ terms plus the Zeeman  energy,
and the transverse part $\hat{\cal H}^\perp$.
The energy of the ferromagnetic state
is always subtracted from $\hat{\cal H}^{zz}$.
A quantum state of $n$ isolated localized magnons is not only
an eigenstate of $\hat{\cal H}$ with $E_n = n(H-3)$, 
but is also an eigenstate of $\hat{\cal H}^{zz}$ and 
$\hat{\cal H}^\perp$ separately with $E_n^{zz} = n(H-2)$ 
and $E_n^\perp = -n$. The idea is to show
that for an arbitrary state
$\langle\hat{\cal H}^{zz}\rangle\geq E_n^{zz}$ and
$\langle\hat{\cal H}^\perp\rangle\geq E_n^\perp$.
For $\hat{\cal H}^{zz}$ the formulated relation holds trivially:
once all spin flips sit on different bonds the Ising
part of the Hamiltonian has the minimal energy $E_n^{zz}$.
In order to show that the same is true for the transverse part
we map the subspace of $n$ spin flips  on the Hilbert space
of $n$ hard-core bosons ${\cal B}_n$. The transverse part 
of the Heisenberg Hamiltonian is, then, the kinetic 
energy of bosons $\hat{\cal H}^\perp\equiv \hat{\cal K}$. 
In addition, we define the Hilbert space of $n$ bosons 
without the hard-core constraint: ${\cal B}_{0n}\supset{\cal B}_n$.
The minimum of the kinetic energy is easily found in 
${\cal B}_{0n}$: all $n$ particles should occupy one of 
the low-energy single-particle states given by Eq.~(\ref{wkag}). 
If the allowed quantum states are restricted again to
${\cal B}_n$, from the variational principle the expectation
value of $\hat{\cal K}$ can only increase: 
$\langle\hat{\cal H}^\perp\rangle \geq 
\min_{{\cal B}_{0n}} \langle \hat{\cal K} \rangle=E_n^\perp$.
The presented arguments not only show that isolated
localized magnons have the lowest energy, but they also allow
to check whether these states are the only lowest energy
states (see below).

For the sawtooth chain the two branches of excitations 
in Eq.~(\ref{wsaw}) are separated by a gap. 
The low-energy scattering states are, therefore,
formed by magnons from the flat branch.
In the two-magnon sector there are 
$\frac{1}{2}(N/2)(N/2+1)$ states constructed from
$N/2$ one-magnon states of the lowest branch.
The number of isolated localized magnons in this sector is 
$\frac{1}{2}(N/2)(N/2-3)$. The difference gives the number
of scattering states, which is equal to $N$. In real space 
representation the above scattering states correspond to
localized magnons occupying the same valley or adjacent valleys.
This representation suggests the following basis in the subspace
of scattering states: 
$|\psi_i\rangle = |\varphi_i\varphi_{i+1}\rangle$
and $|\widetilde{\psi}_i\rangle = |\varphi_i^2\rangle$, where
the exclusion principle of spin-flips and proper normalization
are implied. These states have nonzero overlaps:
\begin{equation}
\langle\psi_i|\psi_{i\pm 1}\rangle = 1/35\,, \ \ 
\langle\psi_i|\widetilde{\psi}_i\rangle = 
\langle\psi_i|\widetilde{\psi}_{i+1}\rangle = 
\sqrt{5/49}\,.
\end{equation}
We calculate the matrix elements of the Hamiltonian between
these states and use the variational principle to determine
the lowest possible energy. A simple variational state 
$|k\rangle = \sum_i e^{-ikr_i}|\psi_i\rangle$ has the energy 
$E(k)=\langle k|\hat{\cal H}|k\rangle/\langle k|k\rangle$.
The minimal value  of $E(k)$ is reached at $k=0$ with the gap
$\Delta=\frac{20}{37}\approx 0.54$, which separates 
the two-magnon scattering states from the low-energy 
boundary $2(H-2)$. An improved variational ansatz $|k\rangle=
\sum_i e^{-ikr_i}(|\psi_i\rangle+c|\widetilde{\psi}_i\rangle)$ 
yields $\Delta \approx 0.44$, which compares well with 
the exact diagonalization result 0.42.\cite{1d} 
The gap is determined mostly by a nearest-neighbor repulsion of 
localized magnons in adjacent valleys. The dispersion of bound 
two-magnon complexes $E(k)$ is weak and does not exceed 10\%.
We conjecture that the repulsion leads to a similar 
behavior in all $n$-magnon sectors: the lowest 
energy states with $E_n=n(H-2)$ are separated from the 
scattering states by a finite gap $\Delta_n=O(1)$. At 
temperatures $T\ll\Delta_n$ one can neglect the higher energy 
states and consider only contribution of isolated localized 
magnons. The latter problem is equivalent to a one-dimensional
lattice gas of particles with energies $(H-2)$ and 
on-site and nearest-neighbor exclusion principle (classical 
hard-dimer model).

The above consideration is straightforwardly extended to
the kagom\'e antiferromagnet. The two-magnon sector consists of 
$\frac{1}{2}(N/3)(N/3+1)$ states constructed from one-magnon 
states of the flat branch $\omega_{1{\bf k}}$. They include
$\frac{1}{2}(N/3)(N/3-7)$ states constructed by placing localized
spin flips on isolated hexagons (called below hard-hexagon states)
and the additional $\frac{4}{3}N$ defect states. The latter states 
correspond to $N$ states of localized magnons, which occupy 
adjacent hexagons, and to $\frac{1}{3}N$ states with two spin 
flips on the same hexagon. We define the real-space basis 
for the defect states as $|\psi_{\alpha i}\rangle 
= |\varphi_i\varphi_{i+\delta_\alpha}\rangle$ and 
$|\widetilde{\psi}_i\rangle = |\varphi_i^2\rangle$, where
$\delta_1=(1,0)$ and 
$\delta_{2,3}=(\frac{1}{2},\pm\frac{\sqrt{3}}{2})$
are three nearest-neighbor sites on a triangular lattice
of hexagons. Using a simple variational ansatz
$|{\bf k}\rangle = \sum_{i\alpha} e^{-i{\bf k}{\bf r}_i} 
c_\alpha |\psi_{\alpha i}\rangle$ we find the gap $\Delta\approx 0.24$
between the isolated localized states and
the scattering defect states.

The low-energy sector of the kagom\'e  antiferromagnet has also two
additional features. First, the hard-hexagon states are not the only
independent localized magnon states. An extra two-magnon state
is illustrated in the left part of Fig.~\ref{geometry}
and consists of two spin flips cycling around a small and a large 
hexagons. Such a state is a special combination of the defect states 
$|\psi_{\alpha i}\rangle$ and $|\widetilde{\psi}_i\rangle$.
Combinatorial arguments show that the extra states give only
$1/N$ corrections to the hard-hexagon result 
in $n$-magnon sectors with small $n$ and disappear completely in 
sectors with $3n/N\agt 0.1$

Second difference with the sawtooth chain is that
the gap for one of the dispersive branches
in Eq.~(\ref{wkag}) also vanishes at $H=H_c$. The 
corresponding propagating magnon has the same energy
$\omega_{3{\bf k}=0}=H-3$ as localized magnons from 
the flat branch. The ground states in the two-magnon 
sector contain apart from the isolated localized magnons
also superposition states of one localized magnon and
one propagating magnon. Such states form a continuum above
the low-energy threshold $E_2=2(H-3)$.
The same is true for all $n$-magnon sectors 
with $n\ll N$. Once $n$ becomes a finite fraction of 
$N$, {\it i.e.}\ for partially magnetized states,
the above picture changes. The low-energy propagating  
magnons experience  multiple scattering from an infinite 
number of localized magnons. Such an interaction produces  
a finite shift of their energy. The exact value of the
energy shift for the $\omega_{3{\bf k}}$ depends on
a precise pattern of localized magnons. It can be explicitly 
estimated for a `uniform' state of $n$-localized magnons, 
corresponding to a low-temperature ensemble of states  
with different translational patterns. We bosonize the 
spin Hamiltonian using, {\it e.g.\/}, the 
Holstein-Primakoff transformations.
The quadratic terms give the excitation
spectrum, which coincides with Eq.~(\ref{wkag}). 
To treat the effect of interactions we define 
two Hartree-Fock averages $m_1=\langle b_i^\dagger b_i\rangle$
and $m_2=\langle b_i^\dagger b_j\rangle$. After
decoupling the four-boson terms the effect of
interaction is reduced to the renormalization
$S\to (S-m_1+m_2)$ in the quadratic terms, where $S=1/2$
is a local spin. At the saturation field $H_c=6S=3$
the gap for the dispersive mode is
$\Delta_d = 6(m_1-m_2)$. For a single localized magnon 
one finds $m_1=-m_2=\frac{1}{6}$.
Therefore, the dispersive mode opens a finite gap
$\Delta_d = 3n/N$. At low temperatures we neglect 
these higher-energy states and reduce the isolated localized 
magnons on small hexagons to a gas of particles on 
a triangular lattice with on-site and nearest-neighbor exclusion,
which is also called a hard-hexagon model. 
A remarkable feature of this model is that it has
an exact solution, which was obtained by Baxter.\cite{baxter}
Below, we discuss the low temperature properties
of the kagom\'e antiferromagnet and the sawtooth chain
in the framework of the two exactly solvable models.

The partition function of the classical hard dimers is
\begin{equation}
Z = \sum_{\{\sigma\}} 
\exp\left[-\frac{\varepsilon}{T}\sum_i\sigma_i\right]
\prod_{\langle ij\rangle} (1-\sigma_i\sigma_j) \ ,
\label{Zsaw}
\end{equation}
where $\varepsilon=H-H_c$ is the chemical potential and 
$\sigma_i=0,1$ are the occupation numbers for dimers
the index $i$ runs over the $N/2$ sites of the basal chain. 
The nearest-neighbor exclusion principle is imposed by 
the last term. The partition function (\ref{Zsaw}) 
can be rewritten in terms of the transfer matrices 
${\cal T}(\sigma_i,\sigma_{i+1})$ as
$Z={\rm Tr}{\cal T}^{N/2}$. In the thermodynamic limit 
$N\to\infty$ the free energy is determined by the 
largest eigenvalue of the transfer matrix:
\begin{equation}
F/N = -\frac{1}{2}\,T
\ln\left(\frac{1}{2} + \sqrt{\frac{1}{4} + e^{-\varepsilon/T}}
\right) .
\end{equation}
From this expression one can calculate various thermodynamic
properties of the sawtooth chain. At $H=H_c$ the entropy and 
the magnetization have universal values independent of 
temperature: 
\begin{equation}
S/N=\frac{1}{2}\ln\left(\frac{1+\sqrt{5}}{2}\right), \ \ \ \  \
M/N = \frac{1+\sqrt{5}}{4\sqrt{5}}\ .
\label{ST0}
\end{equation}
The entropy as a function of field reaches a sharp maximum at $H=H_c$, 
which amounts to 34.7\% of the total entropy $N\ln 2$ of the sawtooth chain.
At low temperatures all the magnetization curves $M(H)$ cross
at $H=H_c$. Comparison between analytical and numerical exact diagonalization 
results shows a nice agreement below $T^*\approx 0.1$.\cite{1d}
The correlation length of the hard dimers is
obtained from the ratio of the two eigenvalues of ${\cal T}$,
and the result is
\begin{equation}
\xi^{-1} = \ln \frac{\sqrt{1+4e^{-\varepsilon/T}}+1}
{\sqrt{1+4e^{-\varepsilon/T}}-1} ,
\label{corr}
\end{equation}
with the wavenumber $\pi$.

\begin{figure}[t]
\begin{center}
\includegraphics[width=0.8\columnwidth]{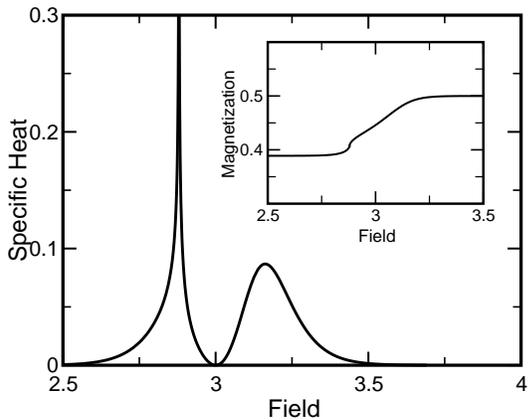}
\end{center}
\caption{\label{Kaghscan} 
Field dependence of the specific heat per one site (in units of $k_B$) 
of a spin-1/2 kagom\'e antiferromagnet at $T = 0.05$
obtained in the hard-hexagon model.
The inset shows the magnetization at the same $T$. }
\end{figure}

The partition function of the hard-hexagon model is given by 
the same expression (\ref{Zsaw}), where the site index runs 
over triangular lattice formed by the centers of hexagon 
voids of the original kagom\'e lattice. This model was solved 
by Baxter, who used a corner transfer matrix method to obtain 
$Z$ as a function of the fugacity 
$z=e^{-\varepsilon/T}$.\cite{baxter} A number of exact 
results following from the Baxter's solution can be transferred 
to the spin-1/2 kagom\'e antiferromagnet using a trivial rescaling. 
The entropy as a function of field at constant $T$ reaches a 
maximum at $H=H_c$. At the saturation field
both the entropy and the magnetization have 
temperature independent universal values:
\begin{equation}
S/N = 0.11108\ , \ \ \ \ \ M/N = 0.44596 \ .
\label{KT0}
\end{equation}
More important results concern a phase transition,
which corresponds to a weak crystallization of 
hard hexagons or localized magnons in the two models:
\begin{eqnarray}
&& H_c(T) = H_c - T\ln z_c\ , \ \ 
z_c = (11+5\sqrt{5})/2 \ , \nonumber\\ 
&& M_c/N = 
{\textstyle \frac{1}{2}}-{\textstyle \frac{1}{3}}\rho_c \ , \ \ 
\rho_c = (5-\sqrt{5})/10 \ .
\end{eqnarray} 
The second order transition at $H_c(T)$ belongs to the universality class
of the three-state Potts model and has the 
exact critical exponents $\alpha=\frac{1}{3}$, $\beta=\frac{1}{9}$,
and $\nu = \frac{5}{6}$. The transition field has a linear temperature
dependence, while the magnetization $M_c$ stays along the critical line.
Variations of the specific heat and the magnetization 
of the spin-1/2 kagom\'e antiferromagnet 
derived from the hard-hexagon model\cite{baxter}
are presented in Fig.~\ref{Kaghscan}.
Baxter used the order parameter defined as a difference
of hexagon densities $\rho_i$ on adjacent sites:
$R=\rho_i-\rho_{i+\delta}$. Up to a renormalization
prefactor $R$ is related 
to the Fourier harmonics $\rho_{\bf q}$ 
at ${\bf q}=(4\pi/3a^*,0)$, where $a^*$ is a period
of the triangular lattice. The magnon crystal has two types
of spins, on hexagons with localized magnons
and between them, which have different average magnetizations:
$\langle S^z\rangle = 1/3$ and 1/2 at $T=0$, respectively.
Hence, the crystalline order of magnons is reflected
in appearance of the corresponding Fourier harmonics in
the structure factor $S^{zz}({\bf q})$ and can be in principle
observed by elastic neutron scattering.

In conclusion, the above analysis 
of the magnetothermodynamics of the quantum kagom\'e antiferromagnet
and of the sawtooth chain model has shown
that strongly correlated spin systems have emergent 
``paramagnetic'' degrees of freedom, which experience an effective
magnetic field $h=H-H_c$. Such effective ``paramagnetic moments''
are responsible for an enhanced magnetocaloric effect
of quantum geometrically frustrated antiferromagnets
in the vicinity of the saturation field 
very similar to their classical counterparts.\cite{mzh03}
During preparation of the manuscript, we have learned
about a recent preprint,\cite{derzhko} where 
the authors have obtained a zero-temperature entropy of the two models
in agreement with our Eqs.~(\ref{ST0}) and (\ref{KT0}),
though they have not discussed
the low-temperature thermodynamics.

We thank G. Jackeli and A. Honecker for useful discussions.
Stay of M.E.Z. at the YITP of Kyoto University was supported 
by a Grant-in-Aid for the 21 century COE, 
``Center for Diversity and Universality in Physics''.
H.T. is supported by a Grant-in-Aid from the Ministry of
Education, Science, Sports, and Culture of Japan.

\end{document}